\documentclass[a4paper]{jpconf}

\usepackage{graphicx}
\usepackage{tikz}
\usepackage{url}
\usepackage{upgreek}
\usetikzlibrary{shapes.geometric}
\usepackage{subcaption}
\usepackage{graphicx} 
\usepackage{changes}  
\usepackage[backend=bibtex, giveninits=true,
style=numeric,sorting=none]{biblatex}
\graphicspath{{figures/}}

\begin{document}
\title{Polarity switch of PMMA powder transported through a PMMA duct}

\author{Wenchao Xu$^1$, Holger Grosshans$^{1,2}$}
\address{$^1$Physikalisch-Technische Bundesanstalt (PTB), Braunschweig, Germany}
\address{$^2$Otto von Guericke University of Magdeburg, Institute of Apparatus- and Environmental Technology, Magdeburg, Germany}
\ead{wenchao.xu@ptb.de} 

\begin{abstract}
During pneumatic conveying, powder electrifies rapidly due to the high flow velocities. In our experiments, the particles even charge if the conveying duct is made of the same material, which might be caused by triboelectrification between two asymmetric contact surfaces. Surprisingly, we found the airflow rate to determine the polarity of the overall powder charge. This study investigates the charging of microscale PMMA particles in turbulent flows passing through a square PMMA duct.
The particles are spherical and monodisperse. A Faraday at the duct outlet measured the total charge of the particles. At low flow velocities, the particles charged negatively after passing through the duct. However, the powder's overall charge switched to a positive polarity when increasing the flow velocity.
\end{abstract}

\section{Introduction}
Triboelectric charging, also known as contact electrification, is the exchange of electric charge between two objects when they touch against each other. The pneumatic conveyance of powders is commonly involved in industrial productions. Being transported pneumatically through ducts or pipelines, powders or particles undergo undesired charging via triboelectrification due to particle--particle and particle--surface interactions. The charge can accumulate on non--conductive materials, leading to a "cone discharge" and igniting dust clouds \cite{glor1989discharges}. 

The transfer of charge in triboelectrification processes typically follows a consistent direction and is determined by uniform global parameters. For example, in the case of different metals, the direction of charge transfer is determined by factors such as the work function \cite{Lowell1975}. For insulators, Lewis basicity is one of many factors that can explain the observed charge transfer behavior \cite{zhang2019Rationalizing}. However, it is important to note that there are cases where the charge or charging direction reverses.

In a study by Tanoue et al. \cite{tanoue2005Polarity}, the polarity change of particles impacting a rotating metal target was investigated. The researchers observed that the tribo-charge polarity depended on the impact angle, resulting in charge reversal under certain conditions. Furthermore, charge reversal has also been observed during sliding contacts \cite{lowell1986Triboelectrification}, \cite{shaw1928Triboelectricity}. These findings highlight that in specific situations, such as impact angles or sliding contacts, the direction of charge transfer can be reversed or exhibit different charging behavior than what is typically observed.

Grosjean and Waitukaitis \cite{grosjean2023Asymmetries} have proposed an explanation for triboelectric charging using the mosaic model, which incorporates donor/acceptor asymmetries. By considering the asymmetrical geometry of the surfaces, they successfully reproduced the observed sign flips in contact sliding experiments. 

\section{Experiment setup and measurement}
\begin{figure}[tb]
	\centering
	\resizebox{0.8\textwidth}{!}{
	    \begin{tikzpicture}[thick]
		\draw [->,>=latex,ultra thick] (0.5,0.25) --node[right,above]{Inlet} (1.5,.25);
		\draw [] (1.5,.0) rectangle (9,.5) node[midway] {Test duct};
		\draw [->,>=latex,ultra thick] (2.0, 1.25) node[right,above]{Powder feeding} -- (2.0, .5);
		\draw [fill=gray!40,ultra thick] (9,-.3) rectangle (11.3,0.8) node[midway] {Faraday cage};
		\draw [ultra thick, dashed] (9,0) -- (11.3,0);
		\draw [ultra thick, dashed] (9,0.5) -- (11.3,0.5);
		\draw [] (11.3,.0) rectangle (12.3,.5) node[midway] {Duct};
		\draw [->,>=latex,ultra thick] (12.3,0.25) --node[right,above]{Air blower} (14,.25);
		\draw [thin] (11.3,.8) -- (11.8,.8);
		\draw [] (11.8,0.8) rectangle (14,2.3) node[align=center,midway] {Charge\\amplifier \& \\electrometer};
		\draw [thin] (1.5,-.2) -- (1.5,-.5);
		\draw [<->,>=latex,thin] (1.5,-.35) -- (2.0,-.35) node[below,midway,align=right] {200 mm};
		\draw [thin] (2,-.2) -- (2,-.5);
		\draw [<->,>=latex,thin] (2.0,-.35) -- (9,-.35) node[below,midway,align=right] {1800 mm};
	    \end{tikzpicture}
	}
	\caption[]{Schematics of the pneumatic conveying system.}
	\label{fig:sketch}
\end{figure}
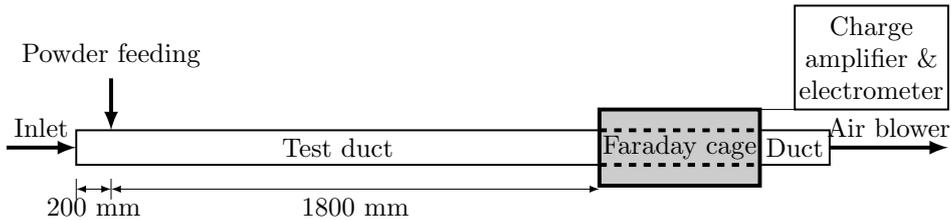

Figure \ref{fig:sketch} illustrates the schematics of our enhanced experimental facility, an upgrade from our previous setup \cite{grosshans2022}. The key components include a square-shaped test duct, an air blower, a powder feeder, and a Faraday cage.

To prevent powder leakage during feeding, the air blower is positioned at the outlet of the test duct to create a negative pressure within the duct. Equipped with a frequency converter, the blower allows for precise control of rotation speed and airflow velocity.
For measuring the bulk flow velocity, we employ a Pitot-tube anemometer. The anemometer is installed in a duct with identical geometry, connected to the air blower's outlet. This setup enables measurement of the streamwise velocity without disturbing the particle-laden airflow.

The test section is a 2-meter-long square-shaped polymethyl methacrylate (PMMA) duct with inner dimensions of 46 mm $\times$ 46 mm, as depicted in fig. \ref{fig:sketch}. 
During each test, we apply a syringe to inject PMMA particles into the test rig through an opening located 200 mm from the inlet of the duct. In the experiments described in this paper, we used monodisperse spherical PMMA particles with two different diameters, namely 100 $\upmu$m and 150 $\upmu$m.

The outlet of the test duct leads to the measurement section, which consists of a PMMA duct enclosed within a cylindrical Faraday cage. Inside the measurement section, a 50 $\mathrm{\upmu m}$ pore-size filter bag is employed to separate particles from the airflow passing through.

To measure the charge of the collected particles, a charge amplifier is connected to the Faraday cage. It is important to note that the electrical circuit experiences consistent charge leakage. However, during the experiment, the time scale for charging measurement was set significantly shorter than the charge leakage time scale of the circuit, so the error due to circuit leakage was negligible.
Furthermore, since triboelectrification is sensitive to ambient conditions, we incorporated a high-accuracy temperature and humidity sensor to monitor and record the temperature and humidity during the experiments.

During the experiment, particles of various diameters were introduced into the test rig, and their resulting charge-to-mass ratios were measured. We obtained the charge-to-mass ratio of particles after their passage through the duct at various air velocities by conducting multiple measurements with different particle sizes and conveying flow velocities.

It is possible that a small number of particles may deposit onto the inner surface of the duct or leak from the duct, thereby evading collection by the filter. To account for this, we measured the weight of particles before and after their introduction into the test rig. The weight difference is carefully controlled to remain below 1$\%$.

\begin{figure}[tb]
\centering
\includegraphics[width=0.55\textwidth]{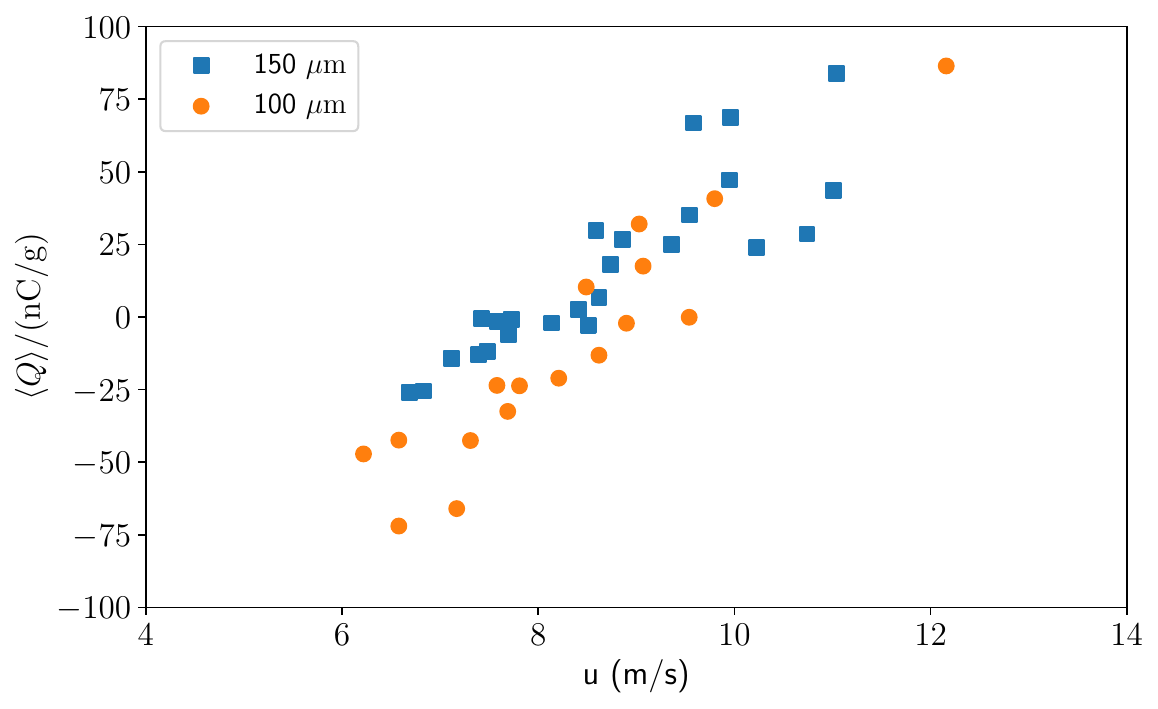}\hspace{2pc}%
\caption{Experiment measurement of charge-to-mass ratio as a function of air velocity at $T = 20.1^\circ C$ and $RH = 58.1\%$.}
\label{fig:q_v_exp}
\end{figure}

Figure \ref{fig:q_v_exp} presents the charge-to-mass ratio of particles after traversing the PMMA duct at various air velocities. We observed a polarity switch in the charging behavior for both particle sizes (100 $\upmu$m and 150 $\upmu$m) at an air velocity range of approximately 8-10 m/s. Specifically, when the air velocity falls below 8 m/s, both particles acquire a negative charge within the PMMA duct. Conversely, they exhibit a positive charge when the air velocity surpasses 9 m/s.

\section{Simulations}
In a recent study by Grosjean and Waitukaitis \cite{grosjean2023Asymmetries}, the switch in charging polarity has been elucidated using "Mosaic models". These models assume that surfaces are composed of a random patchwork of microscopic donor/acceptor sites, leading to persistent global differences even among surfaces made from the same material. Building upon this framework, they expanded the mosaic model by incorporating variations in the densities of donor/acceptor sites across the surfaces.

\begin{figure}[tb]
    \begin{subfigure}{0.35\linewidth}
        \centering
        \includegraphics[trim=0mm 0mm 170mm 0mm,clip=true,width=1\textwidth]{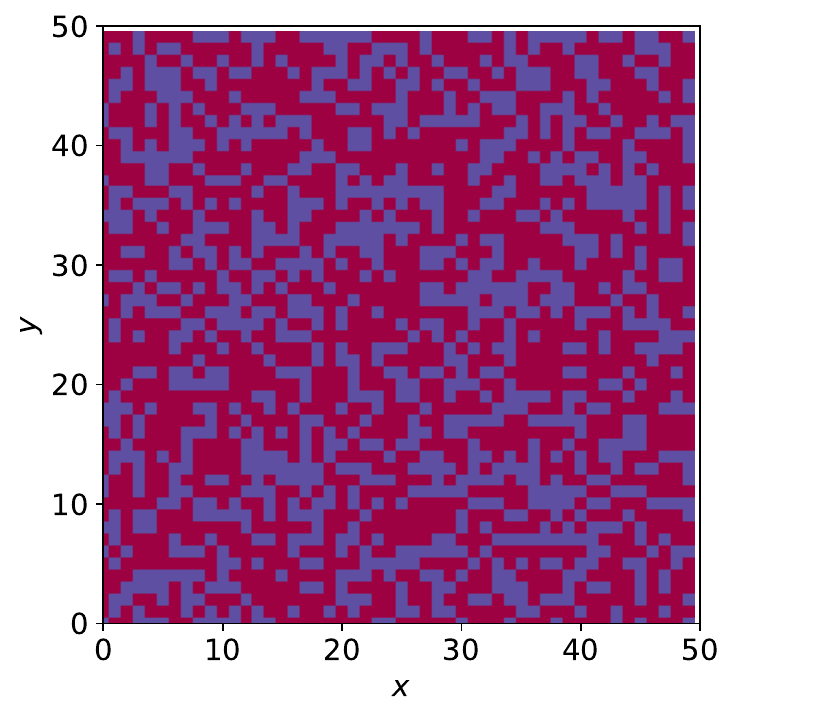}
        \caption{}
        \label{fig:surf_a}
    \end{subfigure}
    \quad
    \begin{subfigure}{0.35\linewidth}
        \centering
        \includegraphics[trim=170mm 0mm 0mm 0mm,clip=true,width=1\textwidth]{figures/surface.pdf}
        \caption{}
        \label{fig:surf_b}
    \end{subfigure}
    \centering
    \caption{A square grid $A$ with $L \times L$ sites contacts on a rectangular grid $B$ with $L \times L$ sites. Red cells represent donor sites and blue cells represent acceptor sites. $L = 50,~p_A = 0.40,~p_B = 0.35$.}
    \label{fig:surf}
\end{figure}

Following their approach, we generated two surfaces, denoted as $A$ and $B$, with dimensions $L \times L$, where $L = 50$. These surfaces are assumed to consist of donor sites with an initial probability of $p_A$ and $p_B$, respectively, while the probabilities for acceptor sites are given by $1-p_A$ and $1-p_B$. To mimic the physics of surface formation, the process also takes into account the favorability of neighboring donor/acceptor sites and includes a time dependency. The transition probability of a given site depends on its neighboring sites and is described by the equations
\begin{equation}
P_A(n_d)=P_0 \exp (-K n_d)~\mathrm{and}~ P_D(n_d)=P_0 \exp (-K(4-n_d)),
\end{equation}
where $P_A$ is the probability of a site transiting into an acceptor and $P_D$ is the opposite. 
Here, $n_d \in [0, 4]$ represents the number of donors in the neighboring grids, and $K$ is a constant that signifies the impact of each neighbor on the local energy barrier \cite{grosjean2020Quantitatively}. The surfaces generated using this approach are illustrated in fig. \ref{fig:surf}. Figure \ref{fig:surf_a} depicts the surface of a particle with $p_A = 0.40$, while fig. \ref{fig:surf_b} represents the inner surface of the PMMA duct with $p_B = 0.35$.

In the simulation, surface $A$ comes into contact with surface $B$ repeatedly to mimic the collisions of particles on the inner duct walls. After each contact, surface $B$ is regenerated randomly with the same parameters. During the contact, donors on surface $A$ transfer a charge of $\mathrm{e}$ to facing acceptors on surface $B$ with the probability $\alpha$ of 0.6, meanwhile acceptors on surface $A$ can also acquire charge from the facing donors on surface $B$ with the same probability.

\begin{figure}[tb]
\centering
\includegraphics[width=0.5\textwidth]{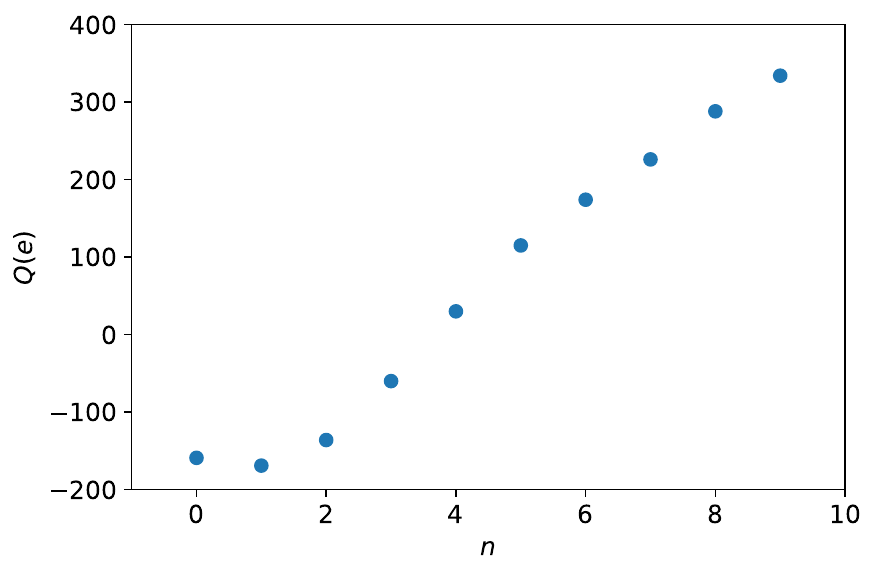}\hspace{2pc}%
\caption{Accumulated charge of the square grid $A$ as a function of contact numbers after repetitively contacting with grid $B$.}
\label{fig:q_n}
\end{figure}

Figure \ref{fig:q_n} illustrates the accumulated charge of grid $A$ as a function of the number of contacts. Notably, the charging polarity undergoes a reversal during the contacting process. Initially, grid $A$ loses electrons in the first few contacts, but subsequently gains electrons in the following contacts. This pattern of electron exchange demonstrates the dynamic charging behavior observed during the contact between surfaces $A$ and $B$.

Turning to the case of particle-laden flows, since we lacked a direct method to measure the frequency of particle-wall contacts, we conducted a series of Large-Eddy simulations (LES) with our open-source CFD tool pafiX \cite{pafix}. The numerical work simulated 100 $\upmu$m particles entrained in turbulent air flow passing through a square-shaped duct with dimensions of 45 $\mathrm{mm} \times 45~\mathrm{mm} \times 2000~\mathrm{mm}$, which corresponds to the meshgrid of $80 \times 80 \times 200$. The boundary conditions in spanwise directions are no-slip walls, whereas in streamwise directions are periodic boundaries.
The fluid is solved in the Eulerian and the particles are solved in the Lagrangian framework using a four-way coupled approach. 
More detailed descriptions of the mathematical model and numerical methods implemented in pafiX were given by Grosshans \textit{et al.}\cite{grosshans2021effect}. 

\begin{figure}[tb]
\centering
\includegraphics[width=0.5\textwidth]{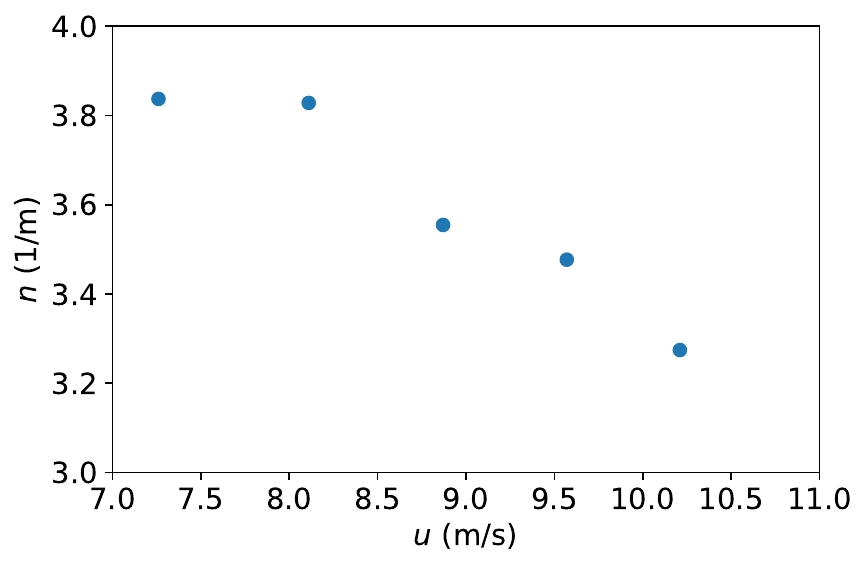}\hspace{2pc}%
\caption{Average wall collision numbers per particle in particle-laden flows as a function of air velocities. The results are from LES simulations with the particle diameter of 100 $\upmu$m. }
\label{fig:sim}
\end{figure}

Figure \ref{fig:sim} displays the average number of particle-wall collisions at various air velocities. Our simulations indicate that air velocity significantly impacts the frequency of particle-wall contacts, highlighting the strong influence of airflow velocity on these interactions, which eventually affects the polarity of the triboelectric charge of the PMMA particles.

\section{Conclusions}
In our experimental investigation of triboelectric charging of PMMA particles in turbulent duct flows, we observed that the polarity of particle charge switches as the airflow velocity changes. Specifically, particles were charged negatively at low velocities and positively at high air velocities.

One explanation for the observed phenomenon is the presence of donor/acceptor asymmetries between the surfaces of particles and the inner walls of the duct. These asymmetries could arise from impurities, contamination, or different surface treatments.
To investigate this hypothesis, we employed the mosaic model and simulated a grid with a higher probability of donors repetitively contacting surfaces with a lower donor probability. We successfully reproduced a scenario where the charge polarity reverses during repeated contact. In this scenario, the polarity of charge is dependent on the number of contacts. Moreover, we conducted a series of Computational Fluid Dynamics (CFD) simulations to study turbulent particle-laden flows. The simulation results strongly indicated that the airflow velocity significantly influences the number of particle-wall contacts, further influencing the particles' charging polarity. 

\ack 
This project has received funding from the European Research Council (ERC) under the European Union Horizon 2020 research and innovation program (grant agreement No. 947606 PowFEct).

\printbibliography
\end{document}